\documentstyle [12pt]{article}

\newcommand{\be}{\begin{equation}}
\newcommand{\ee}{\end{equation}}
\newcommand{\ba}{\begin{array}}
\newcommand{\ea}{\end{array}}
\title{Construction of Simple $q$-Deformed Algebras by Statistics}
\date{}
\begin{document}
\maketitle
\vspace{1cm}
\begin{center}
{\Large S. U.  Park}
\end{center}
\begin{center}
\vspace{.5cm}
Department of Physics,  Jeonju University, \\
1200 Hyoja-3 Chonju, Chonbuk, {\bf 560-759, Korea}
\end{center}
\vspace{1cm}
\begin{abstract}
The simple algebras of a dressed operator, which is composed of a dressing
and a residual operators, are averaged following a proper statistics of
the dressing one.
In the Bose-Einstein statistics, a (fermionic) Calogero-Vasiliev oscillator,
$q$-boson (fermion), and (fermionic) $su_q(1,1)$ are obtained
for each bosonic (fermionic) residual operator.
In the Fermi-Dirac statistics, new similar algebras are derived
for each residual operator.
Constructions of dual $q$-algebras, such as a dual Calogero-Vaisiliev
oscillator,
a dual $q$-boson and a $su_q(2)$, and prospects are discussed.
\end{abstract}
\vspace{1cm}

\section{Introduction}
Quantum groups ($q$-deformation) had been developed in the forms of
deformation of a Lie-Poisson structure \cite{dri,jim}, of a pseudogroup
\cite{wor},
and by using a filtration \cite{rin}.
The quantum groups and their algebras have been studied much in various points
of views in the physics and the mathematics: references and reviews in
\cite{tak}.
any physicists have concentrated on the simplest $q$-algebra known as
$q$-boson algebra (or $q$-Heisenberg-Weyl, $q$-oscillator)
\cite{mac,bie,sun,cha1}.
Its undeformed algebra is a physical canonical commutator,
and other quantum algebras are easily derived from it by the Schwinger method
and the Holstein-Primakoff transformation \cite{lia,cha2,kat,kun}.

The $q$-boson algebra also can be constructed by using
an averaging method \cite{par1,par2} that is different from the methods in
\cite{dri,jim,wor,rin} but similar to the framing in \cite{wit}.
In brief, the construction of $q$-boson consists of three steps.
In first, we construct a contact metric structure
of a Heisenberg group manifold
which generalizes the symplectic (or Lie-Poisson) structure to the odd manifold
of the form of a $U(1)$ principal bundle \cite{arn}.
In second, we extend it into an associated line bundle,
twist by using its fibre, and obtain a canonical action.
Finally, we average the canonical commutator of the extended and
twisted operator ({\em in short, dressed operator}) by using
a photonic partition function (Bose-Einstein statistics), and obtain
the $q$-boson algebra in an effective form.
Extensions to other quantum algebras are still open.

In this paper, I will restrict only on the third step, called
{\em an averaging method}, to see some possibility of extending to other
quantum algebras and statistics (Bose-Einstein and Fermi-Dirac).
In short, the dressed operator is composed with a dressing and a residual
operators like in the Schwinger method,
and the algebra of the dressed operator is averaged following
a proper statistics of the dressing operator.

As results, in the Bose-Einstein statistics,
a (fermionic) Calogero-Vasiliev oscillator,
$q$-boson (fermion), and (fermionic) $su_q(1,1)$ are obtained
for each bosonic (fermionic) residual operator.
In the Fermi-Dirac statistics, new similar algebras are derived
for each residual operator.
Such construction of {\em dual} algebras, such as a dual Calogero-Vasiliev
oscillator,
a dual $q$-boson and a $su_q(2)$, will be derived.

This paper is arranged as follows.
In the next section, I will treat about construction of $q$-deformed algebras
in the Bose-Einstein statistics.
In the third section, I will treat such constructions
in the Fermi-Dirac statistics.
In the fourth section, I will discuss about the dual algebras,
extensions and questions.

\section{Simple $q$-deformed algebras and Bose-Einstein statistics}
\smallskip
We consider a set of
two independent bosons $(a,a^\dagger)$ and $(b,b^\dagger)$\footnote{Hereafter,
in most cases, small (large) characters are used in order to describe
a normal ($q$-deformed) system.}
which satisfy the following commutators;
\be
[~a~,~a^\dagger~]~=~1~=~[~b~,~b^\dagger~].
\ee
The Hilbert space for number operators $(m_a~=~a^\dagger a,~n_b~=~b^\dagger b)$
is spanned by pure states $\mid m_a, n_b \rangle~=~\mid m_a \rangle \mid n_b
\rangle ,
 ~~~m_a,~n_b~=~0,1,2,\cdots$.
In the averaging method, the set $( a , a^\dagger )$ represents
the fibre variables of the line bundle of a Heisenberg group manifold,
and will be averaged out under proper statistics.

The $a$-type boson is assumed to satisfy the Bose-Einstein statistics of which
the partition function for the Hamiltonian $h_a~=~\epsilon_a a^\dagger a$ is
given
\begin{eqnarray*}
Z(\beta) &=& Tr e^{-\beta h_a}, \\
 &=& \sum_{m_a=0}^{\infty} \langle m_a \mid e^{-\beta h_a} \mid m_a \rangle, \\
 &=& \sum_{m_a=0}^{\infty}  e^{-\beta \epsilon_a m_a}, \\
 &=& \frac{1}{1~-~e^{-\beta \epsilon_a}},
\end{eqnarray}
where $\epsilon_a$ is an energy level.
Corresponding to the partition function, we present some expectation values for
later use.
\begin{eqnarray}
\langle ~\theta(m_a+1/2-\alpha)~\rangle &=& \frac{e^{-\alpha \beta \epsilon_a}}
                 {1~-~e^{-\beta \epsilon_a}}, \\
\langle~a a^\dagger~\rangle &=& \frac{1}{(1~-~e^{-\beta \epsilon})^2}, \\
\langle~a^\dagger a~\rangle &=&
                      \frac{e^{-\beta \epsilon}}{(1~-~e^{-\beta \epsilon})^2},
\end{eqnarray}
where $\theta(x)$ is a step function.
In (3), when $\alpha~=~0$, we can recover the partition function (2).
Thus $\theta(m+1/2)$ is equivalent to the unit in the averaging.
In most cases, the averaging is defined on a normalized partition function,
but our unnormalized one does not change things going on.

\subsection{Bosonic $q$-deformed algebras}
\smallskip
Let's start from a two-mode transition \cite{mad} and its operator realization
in the form of a dressed operator.
A two-mode transition, in which a state $\mid m_a,n_b \rangle$ changes
its quantum numbers one level each, is given as
\be
\mid m_a , n_b \rangle ~\rightleftharpoons~~ \mid m_a+1, n_b+1 \rangle.
\ee
The general form of transition may be realized in an algebra by  introducing
new bosonic creation and annihilation operators $(D^\dagger,D)$,
called the dressed operators,
and they are assumed to satisfy the following commutation relations.
\be
[~D~,~D^\dagger~] ~=~ \Theta_{BB}(m_a,n_b),  \label{con1}
\ee
\be
[~m_a~,~D^\dagger~] ~=~ D^\dagger~=~[~n_b~,~D^\dagger~]. \label{con2}
\ee
The commutators (8) show that the dressed operator changes the quantum numbers
$m_a$ and $n_b$ a step each.
The function $\Theta_{BB}(m_a,n_b)$ will depend on the physical situation of
the transition.
In order the dressed operator to realize the transition algebraically,
a Jacobi's identity should be satisfied.
The identity is easily checked for an arbitrary function
$\Theta_{BB}~=~\Theta_{BB}(m_a,n_b)$.
Also to be invariant under the adjoint operation, the $\Theta_{BB}$ should be
real.
For exposition, when $\Theta_{BB}~=~n_b+m_a+1$,
the dressed operators are given as
$D~=~a b,~D^\dagger~=~a^\dagger b^\dagger$.
The set $(D, D^\dagger, \Theta_{BB})$ forms an $su(1,1)$ algebra.
The dressed operators, given by (7) and (8), will
describe algebraically the two-mode transition and more.

Finding a detailed form of the dressed operator is not the ends of this paper,
instead we only need to separate physically it into $(a,b)$-related two parts
to proceed to the averaging method.
For this purpose, I envision a following situation in an atomic and
a solid-state systems \cite{mad,lad}.
The energy level of one mode in the two-mode transition is lying comparatively
close together than the other.
The transition energy spectrum exhibits fine-splitting.
Resulting physical spectrum depends only on the value that averages out
the fine splitting part.
Thus we need to decompose the dressed operator into a fine splitting and
a residual parts.
The fine splitting part is represented by the set $(a , a^\dagger )$,
and the residual part is denoted by new operators $B^\dagger$ and $B$.

The dressed operators $(D^\dagger, D)$ are assumed to be the following forms.
\be
D^\dagger~=~a^\dagger  B^\dagger,~~~D~=~a B,
\ee
This shows that the dressed operator is composed of a dressing operator
$(a,a^\dagger)$ and a residual operator $(B,B^\dagger)$.
Then, from (\ref{con2}) and (9),
the $B^\dagger$ has the property of the $b^\dagger$ boson only,
\be
[~m_a~,~B^\dagger~]~=~0, ~~~[~n_b~,~B^\dagger~]~=~B^\dagger. \label{num}
\ee
In algebraic sense, the set $(B^\dagger,B)$ should satisfy Jacobi's identity
with the other operators.
As a simple solution of the Jacobi's identity, the set $(B^\dagger,~B)$
commute with the set $(a^\dagger,a)$.
The eq. (\ref{con1}) can be rewritten as
\be
a a^\dagger B B^\dagger~-~a^\dagger a B^\dagger B
 ~=~ \Theta_{BB}(m_a,n_b).\label{bb1}
\ee
{}From mutual commuting property of the set $(B^\dagger,B)$ and
$(a^\dagger,a)$,
we can take an expectation value for the state of $a$-type boson
on eq. (11) in order to average out the fine splitting (dressing) part.
By using (2), (4) and (5), we obtain
\be
B B^\dagger~-~q^{-2} B^\dagger B~=~(1~-~q^{-2})^2~
\Theta_{BB}(-\frac{1}{2}q\frac{\partial}{\partial q},n_b)
(\frac{1}{1~-~q^{-2}}),
\label{bb2}
\ee
\be
q^2~=~e^{\beta \epsilon_a}.
\ee
Here we define (obtain) the explicit form of the $q$-parameter related with
the energy level of the fine splitting.
Note that we use the same symbol $(B, B^\dagger)$ for the after-averaged and
before-averaged operators.
We will also take such notations in the other cases.

Let's consider simple physical systems of (\ref{bb2}).

\smallskip
\noindent{\bf Case BB1~)} $\Theta_{BB}(m_a,n_b)~=~m_a~+~n_b~+~1$

\smallskip\noindent
In this choice, the dressed operators $(D,D^\dagger)$ are
generators of a $su(1,1)$ algebra.
Before averaging, the number operator of the set $(B,~B^\dagger)$ is
equal to that of a normal boson.
For consistency, it should contain such solution after averaging.
The eq. (12) shows that the difference from a normal boson becomes
\be
(B B^\dagger~-~b b^\dagger)~-~q^{-2} (B^\dagger B~-~b^\dagger b)~=~0.
\ee
A simple $q$-independent solution of the equation (14) is that $(B^\dagger,B)$
is equal to a normal boson.
\[ B^\dagger B~=~b^\dagger b,~~~~ [~B~,~B^\dagger~]~=~1 \]
After some estimation, we can find a $q$-dependent solution by introducing
a central operator $K$, which acts like a parity operator, such that
\be
K~=~b b^\dagger~-~B B^\dagger~=~(-1)^{n_b},
\ee
\be
{[}~K~,~B^\dagger B~]~=~0~=[~K~,~b^\dagger b~].
\ee
The Calogero-Vasiliev oscillator \cite{vas} is obtained.
\be
[~B~,~B^\dagger~]~=~1~+~2\nu K,
\ee
\be
2\nu~=~q^2~-~1.
\ee

\smallskip
\noindent{\bf Case BB2~)} $\Theta_{BB}~=~1$

\smallskip\noindent
In the case, the dressed operator $(D,D^\dagger)$
itself forms a canonical commutator,
so we can construct the corresponding canonical action \cite{par1,par2}.
{}From (12) and (10), the $q$-boson algebra in \cite{mac} is obtained.
\be
q^2~B B^\dagger~-~B^\dagger B~=~q^2-1, \label{qb}
\ee
\[
{[}~N_b~,~B^\dagger~]~=~B^\dagger,
\]
where $N_b~=~n_b~=~b^\dagger b$.

\smallskip
\noindent{\bf Case BB3~)} $\Theta_{BB}~=~\theta(m_a+1/2)
 ~-~\theta(m_a+1/2~-~2 (n_b+\sigma))$

\smallskip\noindent
The $\Theta_{BB}$ is the function which gets the
unit value on a finite interval
$0~\leq~m_a$ $\leq~2 (n_b+\sigma)$.
{}From (12) and (10) with help of (3), the $su_q(1,1)$ algebra is obtained
\be
q^2~B B^\dagger~-~B^\dagger B~=~(q^2~-~1)(1~-~q^{-4(n_b+\sigma)}).
\ee
In the limit $\sigma \rightarrow \infty$, the function $\Theta_{BB}$ becomes
$\theta(m+1/2)$ which is equivalent to the unit function in the process of
the averaging, thus this system  goes back to the $q$-boson (\ref{qb})
as in \cite{cha}.
An ordinary known $su_q(1,1)$ is obtained by taking transformations
to the generators $J_+,~J_-,~J_0$.
\be
B^\dagger~=~ (q^2~-~1)~q^{-(n_b+\sigma+1/2)}~J_+,~~~~J_0~=~n_b~+~\sigma,
\ee
\be
[~J_0~,~J_\pm~]~=~\pm J_\pm, ~~~
{[}~J_+~,~J_-~]~=~-~[2J_0],
\label{su11}
\ee
where $[x]~=~(q^x~-~q^{-x})/(q~-~q^{-1})$.
This construction with an external parameter $\sigma$
is similar to the $q$-Holstein-Primakoff transformation
in \cite{kat,kun}.

\subsection{Fermionic $q$-deformed algebras}
\smallskip
For a fermion $(f,f^\dagger)$ satisfying an anti-commutator
\[
\{~f~,~f^\dagger~\}~=~1,
\]
we consider a fermionic transition
\be
\mid m_a , n_f \rangle ~~\rightleftharpoons~~ \mid m_a+1, n_f+1 \rangle
\ee
where the fermion number $n_f~=~f^\dagger f$.
Similarly as in the two-mode transition, the transition is assumed to be
governed by fermionic operators with their dressing part
$(a, a^\dagger)$ and the residual fermionic part $(F,F^\dagger)$,
\be
{\cal D}^\dagger~=~a^\dagger  F^\dagger,~~~{\cal D}~=~a F.
\ee
The $F^\dagger$ has the property of the $f^\dagger$ fermion only,
\be
[~m_a~,~F^\dagger~]~=~0, ~~~[~n_f~,~F^\dagger~]~=~F^\dagger.
\ee
The transition, as in the two-mode transition,
can be written as an anti-commutator with a real $\Theta_{BF}$.
\be
\{~ {\cal D}~,~{\cal D}^\dagger~ \} ~=~a a^\dagger F F^\dagger~+~a^\dagger a
F^\dagger F
 ~=~ \Theta_{BF}(m_a,n_f).
\ee
By using (2), (4), and (5), we obtain
\be
F F^\dagger~+~ q^{-2} F^\dagger F~=~(1~-~q^{-2})^2
\Theta_{BF}(-\frac{1}{2}q\frac{\partial}{\partial q},n_f)
(\frac{1}{1~-~q^{-2}}),
\ee

Let's consider about simple physical systems of (27).

\smallskip
\noindent{\bf Case BF1)} $\Theta_{BF}(m_a,n_f)~=~m_a~-~n_f~+~1$

\smallskip\noindent
Before averaging,  the set $(F,~F^\dagger)$ is a normal fermion.
But after averaging, the difference from a normal fermion becomes
\be
 (F F^\dagger~-~f f^\dagger)~+~q^{-2} (F^\dagger F~-~f^\dagger f)~=~0.
\ee
A $q$-independent solution shows that the set $(F,~F^\dagger)$ becomes
a normal fermion.
A $q$-dependent solution of (27) is the fermionic Calogero-Vasiliev oscillator
with a central operator $K_F$,
\be
K_F~=~f^\dagger f~-~F^\dagger F~=~(-1)^{n_f},
\ee
\be
F F^\dagger~+~F^\dagger F~=~1~+~2\nu K_F,
\ee
where $2 \nu~=~q^2-1$ which is equal to that of the BB1 case.

\smallskip
\noindent{\bf Case BF2)} $\Theta_{BF}~=~1$

\smallskip\noindent
The $q$-fermion algebra is obtained.
\be
q^2~F F^\dagger~+~F^\dagger F~=~q^2-1,
\ee
\[
{[}~N_f~,~F^\dagger~]~=~F^\dagger,
\]
where $N_f~=~n_f~=~f^\dagger f$.

\smallskip
\noindent{\bf Case BF3)} $\Theta_{BF}~=~\theta(m_a+1/2)
 ~-~\theta(m_a+1/2~-~2 (n_f+\sigma))$

\smallskip\noindent
The fermionic $su_q(1,1)$ algebra is obtained
\be
q^2~F F^\dagger~+~F^\dagger F~=~(q^2~-~1)(1~-~q^{-4(n_f+\sigma)}).
\ee
In the limit $\sigma \rightarrow \infty$,
it go to the $q$-fermion (31).
This is a fermionic example to obtain a $q$-boson in the limit as in
\cite{cha}.
A fermionic $su_q(1,1)$ is obtained by taking transformations to
${\cal J}_+,~{\cal J}_-,~{\cal J}_0$ and with (25).
\be
F^\dagger~=~ (q^2~-~1)~q^{-(n_f+\sigma+1/2)}~{\cal J}_+,
 ~~~~{\cal J}_0~=~n_f~+~\sigma,
\ee
\be
[~{\cal J}_0~,~{\cal J}_\pm~]~=~\pm {\cal J}_\pm, ~~~
\{~{\cal J}_+~,~{\cal J}_-~\}~=~-~[2{\cal J}_0],
\ee
where $ \{~,~ \}$ is an anti-commutator.

\section{Simple $q$-deformed algebras and Fermi-Dirac statistics}
\smallskip
As we introduce the fine-splitting part in a two-mode transition as
the $a$-type boson, we take the fermionic set $(c,c^\dagger)$, with
an anti-commutator
\[
\{~c~,~c^\dagger~ \}~=~1,
\]
as the fine-splitting part in a fermion-dressed transition.
The eigenstates of the number operator $m_c~=~c^\dagger c$ and
$n_f~=~f^\dagger f$ are $\mid m_c \rangle$, $\mid n_f \rangle$
with respective eigenvalues $m_c~,~n_f~=~0,1$.
The $c$-type fermion is assumed to obey Fermi-Dirac statistics of
a partition function
\begin{eqnarray*}
Z_c &=& Tr e^{-\beta h_c}, \\
    &=& 1~+~e^{-\beta \epsilon_c},
\end{eqnarray}
where the Hamiltonian $h_c~=~\epsilon_c c^\dagger c$ of its
energy level $\epsilon_c$.

We exhibit some formula for later calculations.
\begin{eqnarray}
\langle ~\theta(m_c+1/2-\alpha)~\rangle &=&
             \delta_{\alpha 0}~(1~+~e^{-\beta \epsilon_c})
             ~+~\delta_{\alpha 1}~ e^{- \beta \epsilon_c}, \\
\langle~c c^\dagger~\rangle &=& 1, \\
\langle~c^\dagger c~\rangle &=& e^{-\beta \epsilon_c},
\end{eqnarray}
where $\theta(x)$ is a step function.

\subsection{ Fermionic $q$-deformed algebras with bosonic property}
\smallskip
Let's consider a fermion-dressed bosonic transition in which the state
$\mid m_c,n_f \rangle$ changes its quantum numbers a step each.
\be
\mid m_c , n_f \rangle ~\rightleftharpoons~~ \mid m_c+1, n_f+1 \rangle
\ee
As we are doing before, the transition operators
are assumed to be described by
new bosonic creation and annihilation operators $(D,D^\dagger)$ with their
dressing contents $(c,c^\dagger)$ and the residual part $(F,F^\dagger)$,
\be
D^\dagger~=~c^\dagger F^\dagger,~~~D~=~cF.
\ee
They are assumed to satisfy the following commutation relations.
\be
[~D~,~D^\dagger~]~=~c c^\dagger F F^\dagger~-~c^\dagger c F^\dagger F
 ~=~ \Theta_{BF}(m_c,n_f),
\ee
\be
{[}~m_c~,~F^\dagger~]~=~0, ~~~[~n_f~,~F^\dagger~]~=~F^\dagger.
\ee
Using the results (35), (37) and (38), the above equation (41) is rewritten as
\be
F F^\dagger~-~q_c^{-2} F^\dagger F
     ~=~q_c^{-2}\Theta_{FB}(1,n_f)~+~\Theta_{BF}(0,n_f),
\ee
\be
q_c^2~=~e^{\beta \epsilon_c}.
\ee
Note that the $q_c$ is different from the $q$-parameter of (13).

Let's consider simple cases of (43).

\smallskip
\noindent{\bf Case FB1)} $\Theta_{FB}(m_c,n_f)~=~1~-~m_c~-~n_f$

\smallskip\noindent
Before averaging,  $(F,~F^\dagger)$ is a normal fermion.
But after averaging, difference from a normal fermion becomes
\be
(F F^\dagger~-~f f^\dagger)~-~q_c^{-2}(F^\dagger F~-~f^\dagger f)~=~0.
\ee
In a $q$-independent solution, the set $(F,F^\dagger)$ is a normal fermion.
But, in a $q$-dependent solution of (45), it is the fermionic Calogero-Vasiliev
oscillator with a central operator $K_F$,
\[
K_F~=~f^\dagger f~-~F^\dagger F~=~(-1)^{n_f},
\]
\be
F F^\dagger~+~F^\dagger F~=~1~+~2\nu_F K_F,
\ee
where $2 \nu_F~=~q_c^2-1$.
This value is different from that of BF1 case (30).

\smallskip
\noindent{\bf Case FB2)} $\Theta_{FB}~=~1$

\smallskip\noindent
We obtain
\be
F F^\dagger~-~q_c^{-2} F^\dagger F~=~q_c^{-2}+1,
\ee
\[
{[}~N_f~,~F^\dagger~]~=~F^\dagger,
\]
where $N_f~=~n_f~=~f^\dagger f$.
This shows, in some sense, $q$-bosonization of a normal fermion
$(f,f^\dagger)$.

\smallskip
\noindent{\bf Case FB3)} $\Theta_{FB}~=~\theta(m_c+1/2)
 ~-~\theta(m_c+1/2~-~ (n_f+\sigma))$

\smallskip\noindent
We obtain
\be
F F^\dagger~-~q_c^{-2} F^\dagger F~=~\delta_{n_f+\sigma,1}.
\ee

\subsection{Bosonic $q$-deformed algebras with fermionic property}

\smallskip
Let's consider a fermion-dressed fermionic transition
\be
\mid m_c , n_b \rangle ~\rightleftharpoons~~ \mid m_c+1, n_b+1 \rangle.
\ee
The transition is assumed to be
governed by the fermion-dressed fermion operators;
\be
{\cal D}^\dagger~=~c^\dagger  B^\dagger,~~~{\cal D}~=~c B,
\ee
The transition commutator can be written as
\be
[~m_c~,~B^\dagger~]~=~0, ~~~[~n_b~,~B^\dagger~]~=~B^\dagger,
\ee
\be
{\{}~ {\cal D}~,~{\cal D}^\dagger~\}
       ~=~c c^\dagger B B^\dagger~+~c^\dagger c B^\dagger B
       ~=~ \Theta_{FF}(m_c,n_b).
\ee
After averaging by using (35), (37) and (38), the equation (52) is changed as
\be
B B^\dagger~+~ q_c^{-2} B^\dagger B~=~
      q_c^{-2} \Theta_{FF}(1,n_b)~+~\Theta_{FF}(0,n_b).
\ee

Let's consider  simple cases.

\smallskip
\noindent{\bf Case FF1)} $\Theta_{FF}(m_c,n_b)~=~m_b~-~n_c~+~1$

\smallskip\noindent
\be
 (B B^\dagger~-~b b^\dagger)~-~q^{-2} (B^\dagger B~-~b^\dagger b)~=~0.
\ee
As a $q$-independent solution of the equation (54), the $(B^\dagger,B)$
is equal to a normal boson.
\[ B^\dagger B~=~b^\dagger b.\]
But as a $q$-dependent solution with a central operator $K$,
it is the Calogero-Vasiliev oscillator of fermionic parameter $\nu_F$.
\[
K~=~b b^\dagger~-~B B^\dagger~=~(-1)^{n_b},
\]
\be
[~B~,~B^\dagger~]~=~1~+~2\nu_F K,
\ee
where $2\nu_F~=~q_c^2~-~1$.

\smallskip
\noindent{\bf Case FF2)} $\Theta_{FF}~=~1$

\smallskip\noindent
We obtain a fermionized $q$-boson
\be
B B^\dagger~+~q_c^{-2} B^\dagger B~=~1~+~q_c^{-2},
\ee
\[
[~N_b~,~B^\dagger~]~=~B^\dagger,
\]
where $N_b~=~n_b~=~b^\dagger b$.

\smallskip
\noindent{\bf Case FF3)} $\Theta_{FF}~=~\theta(m_c+1/2)
 ~-~\theta(m_c+1/2~-~(n_b+\sigma))$

\smallskip\noindent
\be
B B^\dagger~+~q^{-2} B^\dagger B~=~\delta_{n_b+\sigma,1}.
\ee

\section{Discussions and Conclusions}
\smallskip
We first discuss about dependence on the choice of operator forms of commutator
in the averaging method.
After averaging, the effective algebra gets little different forms
depending on the choice of operator forms.
As an example, we consider the case BB1 of the two-mode transition
$~[~D~,~D^\dagger~]~=~\Theta_{BB}$, where
\[
D~=~a B,~~D^\dagger~=~a^\dagger B^\dagger,
\]
\[
\Theta_{BB}~=~a^\dagger a~+~b^\dagger b~+~1.
\]
After averaging, we obtain two different solutions; a normal boson and
a Calogero-Vasiliev oscillator as discussed in section 2.1.
If we make a little different choice,
\[
\Theta'_{BB}~=~a^\dagger a~+~ B^\dagger B~+~1,
\]
then, after averaging, the set $(B,B^\dagger)$ becomes only a normal boson
without a path to a Calogero-Vasiliev oscillator.
The fact is same for another choice
\[
\Theta''_{BB}~=~a^\dagger a~+~B B^\dagger.
\]
Following the choice of forms that are explicitly used in the commutator,
we can treat the systems as different ones.
Thus the dependency on the choice of operator forms in an effective algebra
is not dubious things in my averaging method.
Rather it will be interesting to find possible
hidden symmetries which depend on the choice of
operator form.

Also, we want to consider more about relations and loss of information
between the before and after averagings.
As an example, in the BB1 case, the Calogero-Vasiliev oscillator is connected
with the $su(1,1)$ algebra \cite{mac2,cho}.
Also, in the BB2 case, it is related with the contact metric structure of the
Heisenberg group manifold \cite{par1,par2}.
Checking algebraic structures of the new $q$-deformed algebras will be
another interesting issues.

We can consider the {\em dual} systems of above treated four transitions.
\be
\mid m_a,n_b \rangle ~\rightleftharpoons~~ \mid m_a-1, n_b+1 \rangle.
\ee
\be
\mid m_a,n_f \rangle ~\rightleftharpoons~~ \mid m_a-1, n_f+1 \rangle.
\ee
\be
\mid m_c,n_f \rangle ~\rightleftharpoons~~ \mid m_c-1, n_f+1 \rangle.
\ee
\be
\mid m_c,n_b \rangle ~\rightleftharpoons~~ \mid m_c-1, n_b+1 \rangle.
\ee
Then we can easily obtain the dual algebras following similar treatments
for each transitions, but it is rather tedious repetitions.
So I will only state the results of the dual two-mode transition (58).
In order to express a dual system, I will pose the {\em bar} on the top
of the operators.

The dual two-mode transition is given by a set of dual dressed operators
$(\bar{D}~=~a^\dagger \bar{B},$ $ {\bar D}^\dagger~=~a {\bar B}^\dagger)$
with a commutation relation for a given real ${\bar\Theta}_{BB}~=$
${\bar\Theta}_{BB}(m_a,n_b)$.
\be
[~ \bar D ~,~{\bar D}^\dagger~]~=~{\bar\Theta}_{BB}(m_a,n_b).
\ee
After averaging by using (2), (4) and (5), we obtain an effective form
\be
q^{-2}~{\bar B} {\bar B}^\dagger~-~ {\bar B}^\dagger {\bar B}
 ~=~(1~-~q^{-2})^2 {\bar\Theta_{BB}}(-\frac{1}{2}q\frac{\partial}{\partial
q},n_b)
(\frac{1}{1~-~q^{-2}}).
\ee
Simple examples of the above equation are given as follows.

\noindent{\bf Case DBB1~)} ${\bar\Theta_{BB}}(m_a,n_b)~=~m_a~-~n_b$

\smallskip\noindent
In this case, the dressed operators $(\bar{D},\bar{D}^\dagger)$ are
the generators of the $su(2)$ algebra.
A simple $q$-independent result is that $({\bar B},{\bar B}^\dagger)$
is a normal boson.
A $q$-dependent result is the Calogero-Vasiliev oscillator (18) with
the new coefficient $\bar\nu$
\be
2\bar\nu ~=~ q^{-2}~-~1.
\ee

\noindent{\bf Case DBB2)} ${\bar\Theta_{BB}}~=~1$

\smallskip\noindent
It becomes a dual $q$-boson
\be
q^{-2}~{\bar B} {\bar B}^\dagger~-~{\bar B}^\dagger {\bar B}~=~1~-~q^{-2}.
\ee

\noindent{\bf Case DBB3)} $\bar\Theta_{BB}~=~\theta(m_a+1/2)
 ~-~\theta(m_a+1/2~-~2 (n_b-\sigma))$

\smallskip\noindent
The $su_q(2)$ algebra is obtained
\be
q^{-2}~{\bar B} {\bar B}^\dagger~-~{\bar B}^\dagger {\bar B}
 ~=~(1~-~q^{-2})(1~-~q^{-4(n_b-\sigma)}),
\ee
\[
{[}~n_b~,~\bar{B}^\dagger~]~=~\bar{B}^\dagger~~~
{[}~n_b~,~\bar{B}~]~=~-\bar{B}.
\]
Also we can do similar jobs on the other transitions.

Finally we want to extend the averaging method into general quantum algebras,
by possible use of full powers of statistics, and to consider their
geometric meanings in a near future.

\smallskip\noindent
{\large\bf Acknowledgements}

\smallskip\noindent
The author thanks to K. H. Cho for his kind discussions and careful readings.


\end{document}